\documentclass[lettersize,journal]{IEEEtran}
\usepackage{amsmath,amsfonts}
\usepackage{algorithmic}
\usepackage{algorithm}
\usepackage{array}
\usepackage[caption=false,font=normalsize,labelfont=sf,textfont=sf]{subfig}
\usepackage{textcomp}
\usepackage{stfloats}
\usepackage{url}
\usepackage{verbatim}
\usepackage{graphicx}
\usepackage{cite}
\usepackage{booktabs}
\usepackage{hyperref}
\usepackage[capitalize]{cleveref}
\usepackage{cancel}

\crefname{section}{Sec.}{Secs.}
\Crefname{section}{Section}{Sections}
\Crefname{table}{Table}{Tables}
\crefname{table}{Tab.}{Tabs.}

\begin{document}

\title{Brain Dialogue Interface (BDI): A User-Friendly fMRI Model for Interactive Brain Decoding}

\author{Heng Huang, Lin Zhao, Zihao Wu, Xiaowei Yu, Jing Zhang, Xintao Hu, Dajiang Zhu, Tianming Liu\textsuperscript{*}
\thanks{Heng Huang is with the College of Mathematical Medicine, Zhejiang Normal University, Jinhua, China. (email: huangheng05@163.com).}
\thanks{Lin Zhao, Zihao Wu and Tianming Liu are with the School of Computing, The University of Georgia, Athens 30602, USA. (e-mail: \{lin.zhao, zihao.wu1, tliu\}@uga.edu).}
\thanks{Xiaowei Yu, Jing Zhang and Dajiang Zhu are with the Department of Computer Science and Engineering, The University of Texas at Arlington, Arlington, TX, USA. (email:\{xxy1302, jxz7537\}@mavs.uta.edu, dajiang.zhu@uta.edu).}
\thanks{Xintao Hu is with the School of Automation, Northwestern Polytechnical University, Xi'an, China. (email: xhu@nwpu.edu.cn).}
\thanks{Corresponding author: Tianming Liu.}
}

\maketitle

\begin{abstract}
Brain decoding techniques are essential for understanding the neurocognitive system. Although numerous methods have been introduced in this field, accurately aligning complex external stimuli with brain activities remains a formidable challenge. To alleviate alignment difficulties, many studies have simplified their models by employing single-task paradigms and establishing direct links between brain/world through classification strategies. Despite improvements in decoding accuracy, this strategy frequently encounters issues with generality when adapting these models to various task paradigms. To address this issue, this study introduces a user-friendly decoding model that enables dynamic communication with the brain, as opposed to the static decoding approaches utilized by traditional studies. The model functions as a brain simulator, allowing for interactive engagement with the brain and enabling the decoding of a subject's experiences through dialogue-like queries. Uniquely, our model is trained in a completely unsupervised and task-free manner. Our experiments demonstrate the feasibility and versatility of our proposed method. Notably, our model demonstrates exceptional capabilities in signal compression, successfully representing the entire brain signal of approximately 185,751 voxels with just 32 signals. Furthermore, we show how our model can integrate seamlessly with multimodal models, thus enhancing the potential for controlling brain decoding through textual or image inputs.
\end{abstract}


\section{INTRODUCTION}
\IEEEPARstart{B}{rain} decoding techniques have become indispensable tools in neuroscience, especially for elucidating the intricacies of the neurocognitive system and advancing brain-computer interface (BCI)\cite{naselaris2011encoding,luo2023brain,lu2021multi}. These techniques aim to interpret the vast and complex data generated by brain activity, facilitating a deeper understanding of how cognitive processes correspond to neural patterns. Traditionally, brain decoding has relied on methods that attempt to establish direct correlations between neural signals and external stimuli\cite{liu2022decoding,rakhimberdina2021natural,norman2006beyond}. However, these approaches often struggle with the alignment of brain activities to stimuli, primarily due to the dynamic and interconnected nature of neural processes\cite{breakspear2017dynamic,haynes2006decoding,raichle2010two}.

Despite the proliferation of various decoding methods, the challenge of accurately mapping the brain's responses to complex external environments persists. Current methods simplify this task by employing single-task paradigms and classification (or regression) strategies to forge a link between the brain and the world\cite{zhang2021naturalistic,yang2022brain,zaidi2019steps}. For instance, in studies of visual stimulus response\cite{norman2006beyond}, researchers presented subjects with a variety of images (such as objects, faces, or landscapes) and record brain activity using techniques like fMRI. During the experiment, the gathered data is processed using a classification algorithm, specifically a support vector machine, which aims to map brain activity patterns to predefined categories based on their distinguishing features. While this has led to improvements in decoding accuracy, such strategies frequently suffer from limited generalizability across different tasks and contexts. Moreover, these models often encounter difficulties in adequately representing the features of external stimuli which impeding their practical applicability over time\cite{kriegeskorte2019interpreting,haxby2020naturalistic}.

In our study, we refer to these classification-based (or regression-based) methods as static decoding models. This designation stems from the fact that the decoding process is consistently performed on a set of unchanging, predefined categories. In response to these limitations, this paper introduces a novel decoding model designed to facilitate more dynamic and intuitive interactions with the brain. Unlike static models that dominate current research, our approach functions like brain simulators which have the ability to engage with the brain in a manner that mimics natural dialogue. 

Figure 1 provides a summary of our decoding framework. Our model was pretrained on high-quality, open-source fMRI data from the Human Connectome Project (HCP)\cite{van2013wu}, which encompasses a wide range of tasks including motor, working memory, emotion, language, and more\cite{elam2021human}. After training, we represent each subject's entire brain activity (signals from approximately 185,751 voxels) with only 32 components. In our study, these components are referred to as ``injection signals", which encapsulate all dynamic properties of a subject. Once initiated with injection signals, the model functions similarly to the subject's actual brain, enabling us to ``chat" with it as if it were real. To start this conversation, we need to construct specific queries tailored to different scenarios, as presented in Figure 1.

In the first scenario, if we know when an event occurred but not its contents, we need to construct temporal queries. These queries are time series that contain information about the precise timing of events (for more details, see the Methods and Results section). After the construction of temporal queries and input into the model, the model will respond by generating brain spatial maps. These maps identify the regions of the brain that were active during the event, thereby indirectly providing insights into the event's characteristics. Conversely, in the second scenario, if we know the contents of an event and seek to determine its timing, we create spatial queries. These queries are spatial maps of brain activity (also detailed in the Results section), after inputting them into the model, the model will generate a time series which will inform us when this event happened.

In cases where information about the timing and details of an event is lacking, two approaches can be considered. The first involves manually creating a series of queries, each forming a hypothesis about the event. By feeding all hypotheses into a model, the one that elicits the strongest response is likely the correct one. In addition to designing queries manually, it is also feasible in our framework to employ multimodal models such as Contrastive Language–Image Pre-training (CLIP)\cite{radford2021learning,ghandi2023deep} to automate the process.  As depicted in Figure 1. With the advanced capabilities of CLIP, the process of brain decoding can be made more transparent, akin to the interaction between a psychologist and a patient.

Unlike traditional static decoding methods, Our model is developed entirely through unsupervised learning, without reliance on any pre-defined tasks or external information about stimuli during the training phase. This task-free approach eliminates many of the constraints associated with traditional methods, offering a more flexible and adaptive framework. We demonstrate the efficacy of our model through extensive experiments, highlighting its capability to compress whole-brain signals and integrate seamlessly with the existing CLIP model. This integration paves the way for advanced BCI applications, where users can potentially control and interact with brain decoding systems using simple textual inputs. Through our efforts, we aim to contribute to the ongoing advancements and open new avenues for future research in brain decoding technology.

\begin{figure*}[htb]
\centering
\includegraphics[width=0.9\textwidth]{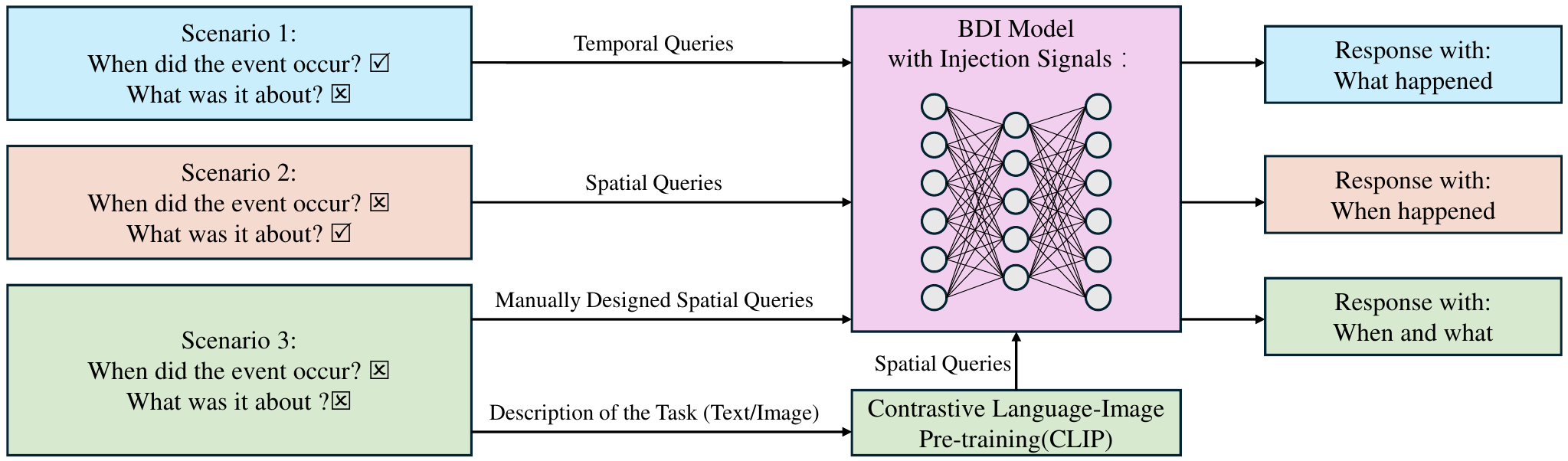}
\caption{Overview of Our Decoding Framework. During the inference phase, the whole-brain activities (fMRI BOLD signals) of a subject we aim to decode are represented by 32 components, referred to as `injection signals' in this study. These injection signals are then used to initialize the model. Once initialized, the model functions analogously to a real brain, enabling us to pose queries and engage in various types of inquiries. Specifically, we consider three scenarios based on whether we possess prior knowledge about the timing or content of events. The complexity of decoding increases with the scarcity of information. For each scenario, we design different types of queries, and the model responds accordingly. In addition to manually designed queries, we also explored the use of CLIP to automate the generation of queries and dialogue using straightforward textual or image inputs. For further details, please refer to Section III.}
\label{fig_1}
\end{figure*}

\section{RELATED WORK}
\subsection{fMRI Brain Decoding Methods}
Brain decoding has captivated researchers for over two decades, resulting in an extensive body of research within this field\cite{awangga2020literature,rybavr2022neural,rogers2023non}. Although our study also focuses on brain decoding, it is crucial to highlight how our approach differs from previous methods. To better outline the current landscape of research, we have broadly classified fMRI decoding techniques into three primary categories: classification-based, regression-based, and reconstruction-based decoding. We will briefly discuss each method in turn and discuss the differences among them.

Firstly, classification-based decoding is a method that categorizes brain activity patterns into distinct groups. A seminal example of this is the study by Haxby et al.\cite{haxby2001distributed}which employed multivariate pattern analysis (MVPA) to classify the category of objects (e.g., faces, houses, chairs) a subject was viewing based on their fMRI brain activation patterns. This approach typically employs classifiers such as Support Vector Machines (SVM), Random Forest, or Linear Discriminant Analysis (LDA), and requires meticulously designed labels for model training \cite{livezey2021deep,brooks2021computational,wagner2019decoding,van2019current}. However, due to the complexity of brain functions, aligning external stimuli with brain activities presents significant challenges. Additionally, the accuracy of aligning brain activity with external stimuli is difficult to quantify and improve.

In contrast, regression-based methods face even more formidable challenges. The objective of this type of method is to predict continuous outcomes from brain activity patterns. For instance, a landmark study by Wager et al.\cite{wager2013fmri} utilized regression analysis to develop a neurological signature capable of predicting the intensity of physical pain. The application of regression decoding models necessitates a profound understanding of the brain's connectivity and the specific cognitive or physiological processes under investigation\cite{sripada2020prediction,you2019fmri,cohen2020regression,kristinsson2021machine}. Additionally, selecting suitable computational algorithms is challenging due to the continuous nature of the training targets.

Reconstruction-based decoding aims to recreate perceptual or cognitive experiences from brain activity data\cite{qiao2022reconstructing,han2019variational,vanrullen2019reconstructing,hoefle2018identifying}. This method has recently attracted considerable attention due to its potential to forge connections between neural activity patterns and the actual content of sensory experiences. For instance, Takagi et al.\cite{takagi2023high} successfully reconstructed high-resolution images from fMRI data using latent diffusion models. Although reconstruction-based methods have significantly advanced our understanding of brain function, like other decoding methods, they rely heavily on precise manual alignment between brain activities and stimuli, such as the images intended for reconstruction\cite{takagi2023high}. Decoding efforts can largely fail if the alignment is not sufficiently accurate.

By examining the challenges of current methods, we recognize that a common factor is their reliance on clearly defined training targets. Specifically, classification-based methods rely on specific categories, regression methods on continuous target variables, and reconstruction methods on sensory content that activates the brain. Based on this characteristic, these methods can be further categorized under what we call ``target-based decoding methods". The specific training targets required by these methods prevent them from disregarding external stimuli or fully utilizing underlying brain functions. To address this issue, our study introduces a ``target-less decoding method." Unlike previous approaches, our goal is to develop a model that does not require a specific target and is not trained for any particular tasks. Instead, our model uses queries to determine the brain state of a subject. These queries serve as the language we use to communicate with the model. Upon receiving a query, the model functions as a digital counterpart of a real brain, retrieving the subject's experiences and responding accordingly. To perform a decoding task, we embed our intentions into these queries and analyze how the model responds. This approach resembles psychological counseling, where a clinician determines a patient's mental state through continuous questioning and awaits feedback.

\subsection{Transformer and Its Applications in fMRI}

In recent years, deep learning\cite{lecun2015deep} has dominated numerous fields, achieving state-of-the-art results across various topics. For instance, convolutional neural networks (CNNs) have surpassed human-level accuracy in visual recognition tasks with natural images\cite{aloysius2017review}. Consequently, an increasing number of researchers have adopted neural networks as their primary methodology in studies. Among these deep learning algorithms, the Transformer\cite{vaswani2017attention} architecture has recently garnered significant attention due to its remarkable effectiveness across a range of tasks in natural language processing, computer vision, and more. Transformer has been particularly adapted for use in general large models, such as the GPT model, and is believed to be instrumental in advancing the capabilities of AI in natural language understanding and generation\cite{wu2023brief}.

For fMRI studies, there also have been numerous successful applications. For instance, Zhao et al. \cite{zhao2022embedding,zhao2023generic,zhao2023coupling} proposed to utilize Transformer to encode the brain function from fMRI into a unified latent space. Yu et al. \cite{yu2023gyri} designed a Twin-Transformer framework to disentangle spatial-temporal patterns of gyri and sulci. Deng et al. \cite{deng2022classifying} utilized spatial-temporal transformers for classifying autism spectrum disorder (ASD). Malkiel et al. \cite{malkiel2022self} proposed a self-supervised transformer for predicting age and gender, and Zhang et al. \cite{zhang2022cnn} employed a CNN-transformer hybrid approach for decoding visual neural activity into text. Caro et al. \cite{ortega2023brainlm} introduced an fMRI foundational model aimed at elucidating the complex spatiotemporal dynamics of human brain activity. These studies highlight the adaptability of transformers to fMRI analysis. In our research, we leveraged the powerful transformer architecture to investigate the target-less decoding approach we previously discussed.

\section{PROPOSED METHOD}
\subsection{Model Architecture and Training}

\begin{figure*}[htb]
\centering
\includegraphics[width=0.9\textwidth]{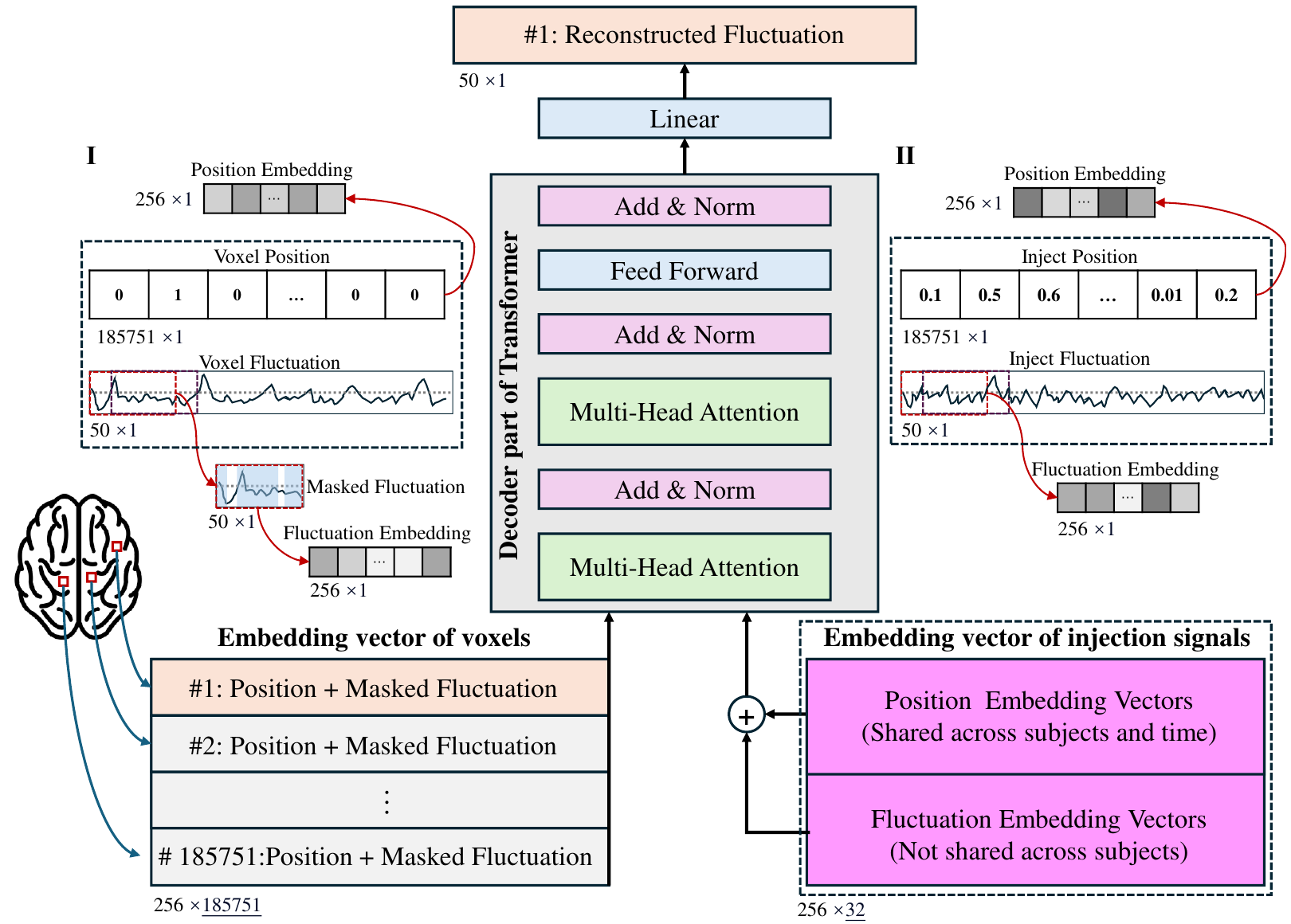}
\caption{Schematic illustration of the model inputs. Part I displays the original form of the voxel signals, detailing both spatial location and temporal fluctuations. Part II shows the injection signals, including both the target brain locations for signal injection and the corresponding signal fluctuations.}
\end{figure*}

Our model architecture is illustrated in Figure 2, with the decoder part of the Transformer as its central component. The computational mechanisms of the Transformer have been extensively explored in previous research\cite{vaswani2017attention,radford2019language,han2022survey,chen2021transformer}. Briefly, the core innovation of the Transformer is its self-attention mechanism, which diverges from earlier models that primarily used recurrent or convolutional layers to handle sequential data. The Transformer uniquely allows the model to dynamically assess and prioritize the relative importance of various elements within a lengthy input sequence. The mechanism that determines the relative importance of different parts of the input is called attention, which can be calculated using the following equation. 

\begin{equation}
Attention(Q,K,V) = softmax(\frac{QK^{T}}{\sqrt{d_K}})V
\end{equation}
Where, the query (Q), keys (K), and values (V) are vectors that are mapped from input elements using a linear layer. The expression $QK^{T}$ represents the dot product between the query and the keys, which helps to measure the compatibility of different elements. A softmax function is then applied to these compatibility scores to obtain weights that are proportional to their importance. The term $\sqrt{d_K}$ is used for scaling, ensuring numerical stability in the computations.

The self-attention of Transformer addresses the memory bottlenecks and gradient issues commonly faced by RNNs\cite{vaswani2017attention}. Furthermore, enables the parallel processing of sequences, greatly benefiting the handling of large models and datasets. Figure 2 illustrates the key components of the Transformer, including multi-head attention, normalization, and feed-forward layers\cite{liu2023gpt}.  In this study, We largely retain the structure of the transformer components, focusing on adapting them to brain activity data (fMRI signals). 

Our model utilizes two types of input. The first comprises fMRI signals recorded from subjects engaged in various tasks. This input encompasses detailed information about the spatial positions of all voxels and their corresponding signal fluctuations. The second type of input is a set of synthesized signals, referred to as ``injection signals." These signals represent a condensed version of the fMRI data and include specific positions and fluctuations. The original forms of fMRI and injection signals are illustrated in Figure 2, parts I and II, respectively. It should be noted that in our study, the position of each voxel (or injection signal) is represented by a one-hot position vector with dimensions of $185,751\times1$, where $185,751$ corresponds to the total number of voxels in the brain. In this setting, we treat each voxel as a fundamental unit for modeling, which enables us to conduct large-scale brain stimulation and fully leverage the high resolution of fMRI data. Moreover, to adapt our model to various tasks, including resting states, we employed a sliding window approach that focuses on a local receptive field of 50 time points on signal fluctuation. The window size was determined experimentally, as detailed in our supplementary materials.

The training objective of our model is to optimize the parameters of the Transformer and identify a set of injection signals that minimize the error in reconstructing masked fMRI inputs. Mathematically, it can be expressed as the following equation.

\begin{equation}
\min_{f_\theta,I_p,I_f}{L(f(V_p,V_{mf},I_p,I_f),{V_f})}
\end{equation}

Where $f$ denotes the Transformer, $f_\theta$ refers to the Transformer's parameters, $V_p$ and $V_{mf}$ represent the position vector and masked fluctuation vector of a voxel respectively. $I_p$ and $I_f$ are the position vectors and fluctuation vectors of injection signals. With ${V_f}'$ being the reconstruction version of $V_f$, the training objective is to optimize $f_\theta$, as well as find  $I_p$ and $I_f$, to minimize the error between ${V_f}'$ and $V_f$.

It should be emphasized that within the transformer, both the position and fluctuation vectors are encoded as embedding vectors and then combined by addition. This step is vital for capturing the spatial and temporal dependencies within the data and has been thoroughly discussed \cite{vaswani2017attention}. In our study, we utilized a shared embedding layer between voxels and injection signals. The corresponding equations are provided below, where, $Embed_p$ and $Embed_f$ represent the embedding layers for the position and fluctuation vectors respectively.

\begin{equation}
V_{embed} = Embed_p(V_p) + Embed_f(V_f) 
\end{equation}
\begin{equation}
I_{embed} = Embed_p(I_p) + Embed_f(I_f) 
\end{equation}

After training the model, the optimized injection signals can be treated as a representational version of the fMRI data. To better elucidate the underlying mechanisms, we have included an illustration in Equation 5, this equation is akin to Equation 2 but explores a more extreme scenario in which only the voxel's position is known, and we seek to recover its fluctuations.

\begin{equation}
\min_{f_\theta,I_p,I_f}{L(f(V_p,I_p,I_f),{V_f})}
\end{equation}

Due to the limited information about voxel fluctuations, the recovery of a voxel relies entirely on the injection signals. The relationship between brain voxels and injection signals can be described by the following attention function.

\begin{equation}
A_{vi} = softmax(Q_{vf} \begin{bmatrix}
K_{i1} &  K_{i2} &  \cdots &K_{in}
\end{bmatrix})
\end{equation}

Where $Q_{vf}$ is the query vector of a voxel, which is projected from the voxel’s position vector. $K_{in}$ represents the key vector of the nth injection signal. $A_{vi}$ is a vector that contains the attention values between the query and the key vectors. For clarity, assume that the transformer model consists of a single attention head and one layer. The fluctuations of the voxel, denoted as $V_f$, can be finally reconstructed using the following equation.

\begin{equation}
V_{f} = A_{vi1} * V_{i1} +  A_{vi2} * V_{i2} +  \cdots +  A_{vin} * V_{in}
\end{equation}

$A_{vin}$ represents the attention value between the voxel and the nth injection signal, while $V_{in}$ is the projected value of the nth injection signal. The equation makes it clear that the injection signal ultimately forms the basis of the voxel, which can be combined to represent voxel fluctuations and allows for the retrieval of specific voxel fluctuations by using its position as a query. It should be noted that in an actual transformer model, the attention values are derived through multiple nonlinear computations. However, this simplification helps illustrate the foundational principles underlying the complex interactions within the model.

Additionally, during model training and inference, an attention mask is applied to establish causality between voxel and injection signals. This ensures that only the injection signals can determine the reconstruction of voxel fluctuations. Conversely, voxel activities do not influence the connections between the injection signals.

\begin{equation}
A_{mask} = \left(
\begin{array}{c|c|c|c|c|c|}
    & V_1 & I_1 & I_2 & \cdots & I_n \\ \hline
V_1 & 1  & 1 & 1&  \cdots  & 1 \\ \hline
I_1 & 0  & 1 & 1&  \cdots  & 1 \\ \hline
I_2 & 0  & 1 & 1&  \cdots  & 1 \\ \hline
\vdots & \vdots  & \vdots & \vdots& \ddots  & \vdots \\ \hline
I_n & 0  & 1 & 1&  \cdots  & 1 \\ 
\end{array}
\right)
\end{equation}

By employing the strategy outlined, we optimized 32 injection signals (experimentally determined, see supplementary materials) to effectively represent the activity patterns of all 185,751 voxels. These signals enable the network to predict fluctuations across voxels in the brain. To enhance the training efficiency of our model, we masked 90\% of the voxel fluctuations, leaving 10\% unmasked during the actual training process. This selective masking achieves an optimal balance between preserving sufficient data and maintaining computational efficiency, details of which are discussed in the results section.

\subsection{Query Construction}
After model training, the inference phase enables us to initiate a dialogue-like conversation with the brain. As illustrated in Figure 4(a), the initial step involves transforming new data into the corresponding injection signal. This process is detailed in Algorithm 1, where $I_p$ and $I_f$ denote the position and fluctuation of injection signals, and $V_p^i$ and $V_f^i$ correspond to those of the nth voxel. During this phase, we freeze the weights of the BDI model and iteratively process all voxels of a subject by feeding their position vectors into the model. Backpropagation is employed to minimize the error between the predicted and actual fluctuations of the voxels. Once completed, the identified injection signals are used to initialize the BDI model for subsequent processes.

\begin{algorithm}[H]
\caption{Find Injection signals for new data}\label{alg:alg1}
\begin{algorithmic}
\STATE $I_p \leftarrow \text{pretrained position}$ \COMMENT{Shared across subjects}
\STATE $I_f \leftarrow \text{random value}$ \COMMENT{Randomly initialized}

\FOR{$V^{i}= 1$ to $n$}
     \STATE $\min_{I_f}{L(f(V_p^i,I_p,I_f,I_p),{V_f^i})}$
\ENDFOR
\RETURN $I_f$ 
\end{algorithmic}
\label{alg1}
\end{algorithm}

As discussed in the Introduction section, three scenarios could occur during our decoding process. For the first scenarios where the timing of events is known and we aim to decode the contents of these events, we need to construct temporal quires to determine the state of the brain. Figure 13(a) illustrates the construction process of the temporal queries. These queries are constructed based on the principle of neurovascular coupling, where event times are convolved with the Hemodynamic Response Function (HRF) to create a time series model that predicts the expected BOLD response. The time series model commonly known as the design matrix has been widely used in General Linear Model (GLM) studies and various other research projects that focus on analyzing the relationship between experimental stimuli and brain activity\cite{poline2012general,beckmann2003general,monti2011statistical,woolrich2004multilevel}. 

To uncover the contents of this event, we need to identify a specific combination of brain voxels that produce a response closely matching the temporal query. While it is possible to manually locate a suitable set of voxels using a brute-force search method, this study employs backpropagation for identification, as illustrated in Figure 4(b). We refer to the set of searched voxels as spatial maps. These maps help pinpoint which brain regions are activated during specific times, thereby providing indirect insights into the contents of the event. For instance, significant activation of the auditory cortex could suggest that the subject is primarily processing sounds at that time.

\begin{figure}[htb]
\centering
\includegraphics[width=\linewidth]{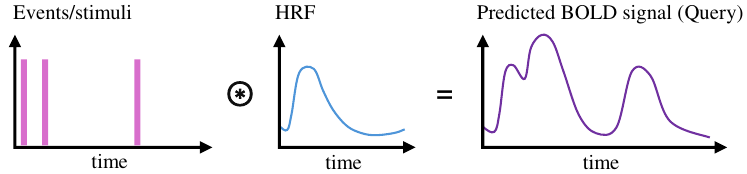}
\caption{Illustration of the construction process for the temporal query. }
\end{figure}

In another scenario, we know the contents of events (specifically, brain activation) and try to decode their exact timing. We can construct a spatial query to elicit a temporal response from the model, as shown in Figure 4(c). The intensity of this response, highlighted with a green arrow, indicates the most likely timing of the event.

The challenge intensifies when both the contents and the timing of the events are unknown, making it difficult to construct either a spatial or temporal query. In such cases, we manually design a sequence of spatial queries as hypotheses by using predefined Regions of Interest (ROIs). These predefined areas act like `words' in a `sentence'—they form the language of the brain, which we use to communicate with it. Further details of this approach will be discussed in the following section.

\begin{figure*}[htb]
\centering
\includegraphics[width=0.9\textwidth]{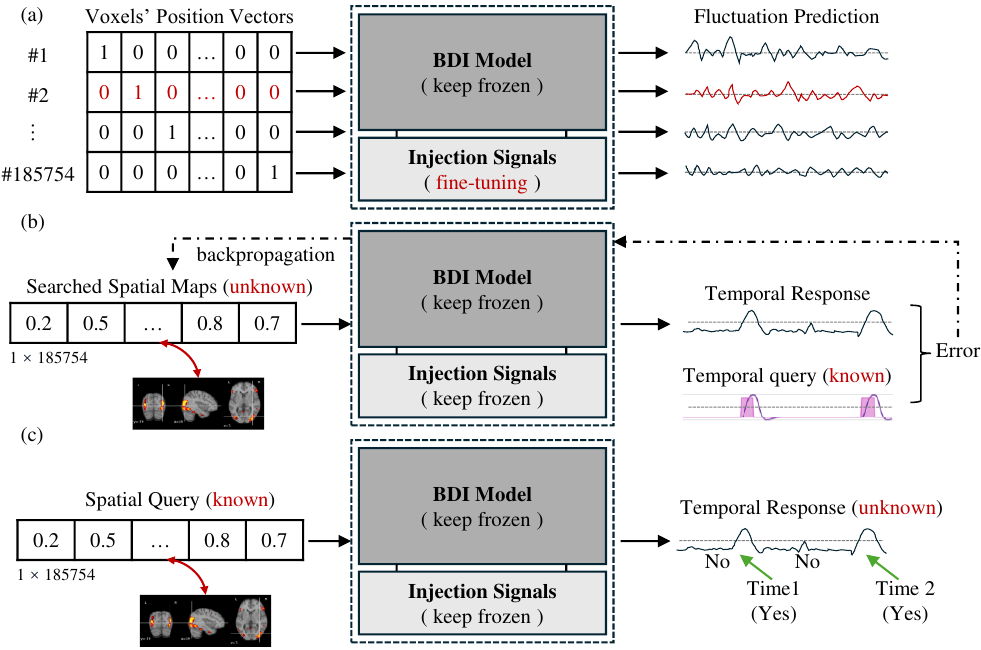}
\caption{Query construction and brain responses in different decoding scenarios. (a) Fine-tuning the model to identify injection signals for a new subject, during which the model's weights are frozen. (b) Temporal query construction when the exact timing of an event is known, but the contents of the event are unknown. (c) Spatial query construction for known event contents when the exact timing is unknown. The spatial query acts like `words' in a sentence, which we use to initiate a dialogue with the brain.}
\end{figure*}

\subsection{Aligning Text and Image with Brain Activity}

\begin{figure*}[htb]
\centering
\includegraphics[width=0.9\textwidth]{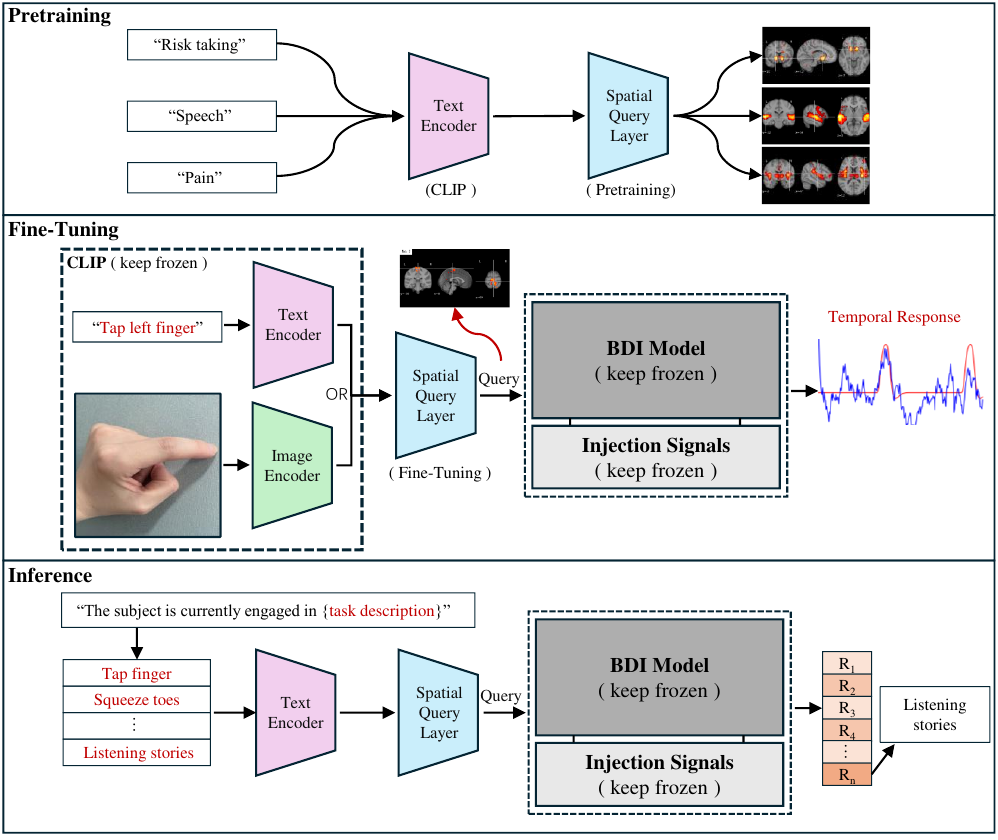}
\caption{Illustration of Aligning Text/Image with Brain Activities for Spatial Query Translation. During the pretraining phase, Neurosynth data, which includes pairs of text and brain maps, is utilized to optimize the parameters of the spatial query layer. In the finetuning phase, the BDI model is integrated as a discriminator to enhance the accuracy of spatial query translation. During evaluation, a series of hypotheses are formulated to assess the brain states of subjects.}
\end{figure*}

Our study also aims to improve interactions with the brain by mimicking human-like dialogue. To achieve this, we need to align human-friendly expressions, such as text or visual images to the spatial queries previously discussed. The integration of the BDI model we propose will facilitate this conversion process.

As illustrated in Figure 5, the core of this alignment involves the CLIP model, a well-known multimodal tool developed by OpenAI\cite{radford2021learning}. CLIP has been pretrained on approximately 400 million image-text pairs, endowing it with the robust and versatile capability to comprehend and generalize across a broad spectrum of images and textual descriptions. As shown in Figure 5 (Middle), CLIP consists of two primary components: a text encoder and an image encoder. These encoders allow us to map text descriptions and images onto high-level feature spaces. This feature space is highly discriminative, enabling direct comparisons by calculating the cosine distance between text and images. To further align text and image with brain activity in our study, we structured our methodology into three parts: meta-analysis data pretraining, BDI model integration for fine-tuning, and model inferencing. We will discuss it one by one.

In the pretraining process, we utilized data from Neurosynth, an open-source dataset specifically designed for the large-scale, automated synthesis of fMRI data\cite{yarkoni2011large}. This dataset includes pairs of text and brain maps, derived through an automated meta-analysis method that incorporates data from over 14,371 studies and features 507,891 activations across various brain locations. For the pretraining, we incorporated a linear layer, termed the `spatial query layer', after the CLIP model. This layer is used to directly map high-level text features into corresponding spatial maps.

The pretraining training objective is to optimize the parameters of the spatial query layer to minimize the discrepancy between text-generated and actual spatial maps from the Neurosynth data, which can be expressed as follows. 

\begin{equation}
\min_{SQL_\theta}{L({Clip}_{text}(D_{text}),D_s)}
\end{equation}

Where $Clip_{text}$ represents the text encoder of CLIP, $D_{text}$ and $D_s$ correspond to the text description and spatial maps from neurosynth data, respectively. $SQL_\theta$ is the parameters of the spatial query layer. Notably, during this pretraining phase, no information from the HCP data task design was used. In subsequent experiments, we demonstrate that this pretraining process could achieve reasonably good alignment, thereby enabling the effective performance of zero-shot classification tasks \cite{xian2018zero}.

After the pretraining is completed, we can further utilize the task design information (event timing) to fine-tune the spatial query layer and enhance the accuracy of the translation. To achieve this, we used the BDI model as a discriminator to assess the correctness of the translation. As expressed in Equation 10, $BDI$ represents our model, and $T_{design}$ refers to the task design of the HCP data. For a successful translation, the BDI model will generate a temporal response that closely matches the task design, also as illustrated in Figure 5 (Fine-Tuning).

\begin{equation}
\min_{SQL_\theta}{L(BDI({Clip}_{text}(D_{text})),T_{design})}
\end{equation}

Finally, in the inference stage, when addressing the third scenario we discussed, we can develop a sequence of hypotheses about the events, Figure 5 (Inference). After converting these hypotheses into text or images, we input them into the model. The model then produces corresponding responses. The hypothesis that triggers the strongest temporal response from the BDI model is likely the correct one.

\section{EXPERIMENTS}
\subsection{Data preparation}
Our study utilized data from the Human Connectome Project (HCP)\cite{van2013wu}, a high-quality, open-source dataset. The HCP data includes eight tasks that cover various aspects of daily human life (Emotion, Gambling, Language, Motor, Relational, Social, Working Memory and REST). We applied standard fMRI preprocessing as detailed in previous studies\cite{glasser2013minimal,huang2017modeling}. Unlike many studies, we treated each brain voxel as an individual sample and used a temporal sliding window of 50. The training sample size for each subject was calculated using the formula $S=\text{number of voxels} \times \text{number of windows for all tasks}$. We employed a gray matter mask to exclude all white matter, resulting in a total of 185,750 voxels, and the number of temporal windows for all tasks is 2,748. Thus, the total number of training samples per subject amounted to 510,443,748. Specifically, the models are trained using 75\% of the dataset, while the remaining 25\% is utilized as the testing set.

\subsection{Impact of Injection Signals on fMRI Reconstruction.}
In this section, we explore the impact of the injection signal on fMRI signal reconstruction. The injection signal comprises two main components: the injection position and its corresponding signal fluctuation. We analyze the effects of these components separately.
Figure 6 (top) illustrates the initial phase of our study, where we conducted two experiments on 0\% hint signal reconstruction using test data. In the first experiment, to examine the effects of injection fluctuations, we kept all injection position vectors constant and gradually reduced the amount of fluctuation by setting them to zero. The test error resulting from this adjustment is illustrated by a solid black line in Figure 6. In contrast, in the second experiment, we maintained the signal fluctuation vectors unchanged while progressively removing the injection position information. The results of this modification are indicated by a red dashed line in Figure 6. For comparison, the dashed blue line in Figure 6 represents the results obtained when the quantity of injection position vectors and fluctuation vectors are reduced simultaneously.

\begin{figure}[htb]
\centering
\includegraphics[width=\linewidth]{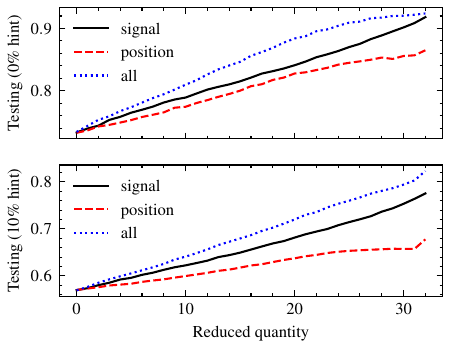}
\caption{Evaluating the Influence of Injection Signals on fMRI Signal Reconstruction. Top: reconstruction error of 0\% hint experiments. Bottom: This demonstrates the impact of introducing a 10\% hint, which leads to a significant decrease in reconstruction error.}
\end{figure}

Our observations indicate that the loss of injection signal information results in a significant rise in testing error, which increases approximately linearly with the degree of information loss. Both the injection position and signal fluctuation are critical factors in the reconstruction of fMRI signals. Furthermore, our analysis suggests that signal fluctuation has a more pronounced effect than the injection position. This is evidenced by the testing error curve for fluctuation loss, which is positioned higher than that for position loss. Remarkably, when all fluctuation information is eliminated, the testing error approaches the level observed when both fluctuation and position information are absent.

After conducting the 0\% hint reconstruction experiments, we performed additional tests where the model was exposed to 10\% of the information from the signals. This exposure corresponds to roughly five time points (10\% of the total length of the input window which consists of 50 time points), as illustrated in Figure 6 (bottom). The results indicate a significant reduction in overall testing error. Specifically, the average testing loss decreased from 0.73 to 0.57, corresponding to an improvement of approximately 21.9\%. This significant improvement demonstrates the effectiveness of incorporating hints to boost signal inference in fMRI signal reconstruction.

Furthermore, to identify which task types our model excels at, we analyzed the testing reconstruction error for different task types, as depicted in Figure 7. Our model generally shows consistent performance across a variety of tasks. However, we observed that certain tasks, specifically the language task (marked as LAN in Figure 7) and the rest task (marked as RST in Figure 7) are more challenging to model. In subsequent experiments, we found that providing the model with hints reduces the variation in performance across tasks. As illustrated in Figure 7 (red bars), offering a 10\% hint significantly decreases the testing error for both the language and resting tasks, with reductions of approximately 21.93\% and 25.99\% respectively.

\begin{figure}[htb]
\centering
\includegraphics[width=\linewidth]{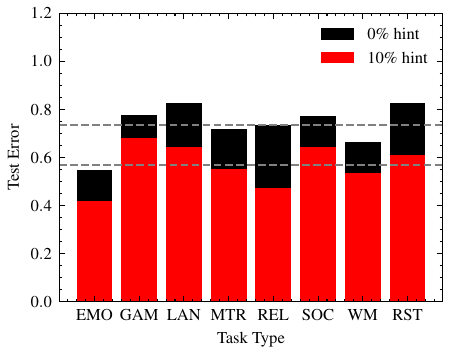}
\caption{The performance of signal reconstruction across various tasks. Including Emotion (EMO), Gambling (GAM), Language (LAN), Motor (MTR), Relational (REL), Social (SOC), Working Memory (WM), and Rest (RST). The black bars represent the reconstruction errors without hints (0\% hint), and the red bars illustrate the enhanced accuracy obtained when the model is provided with a 10\% hint.}
\end{figure}

The efficacy of deep neural networks is often credited to their remarkable data compression capabilities. It is believed that a model's ability to compress data effectively signifies its potential to reveal the underlying structure of the input data. Although this hypothesis requires further investigation and the inner workings of deep neural networks remain mysterious, our experiment offers some preliminary insights. We observed notable data compression performance in our model. In our study, the brain activity of each subject, captured by 185,751 voxels, is efficiently summarized into 32 injection signals. This results in a compression ratio of approximately 5804.7, demonstrating the model's capability to distill the dynamic activity of the entire brain into a simplified representation.

\subsection{Injection Signal Analysis}
We have examined the performance of fMRI signal reconstruction using our model. To gain deeper insights, we sought to address several pertinent questions regarding the nature of the injection signal.

To address our primary question, we analyzed all 32 injection position vectors identified through our research. The visualization of these positions in brain space is presented in Figure 8 where each subfigure represents an ``injection map" with values ranging from 0 to 1. These values provide a statistical perspective: a value near 0 suggests a lower likelihood of signal injection at that location, whereas a value closer to 1 indicates a higher probability. Figure 8 shows that the distribution of injection maps is not random but exhibits a clustered pattern, akin to conventional spatial patterns observed in fMRI studies, suggesting that each injection is linked to specific, potentially multiple, functional brain areas.

\begin{figure}[htb]
\centering
\includegraphics[width=\linewidth]{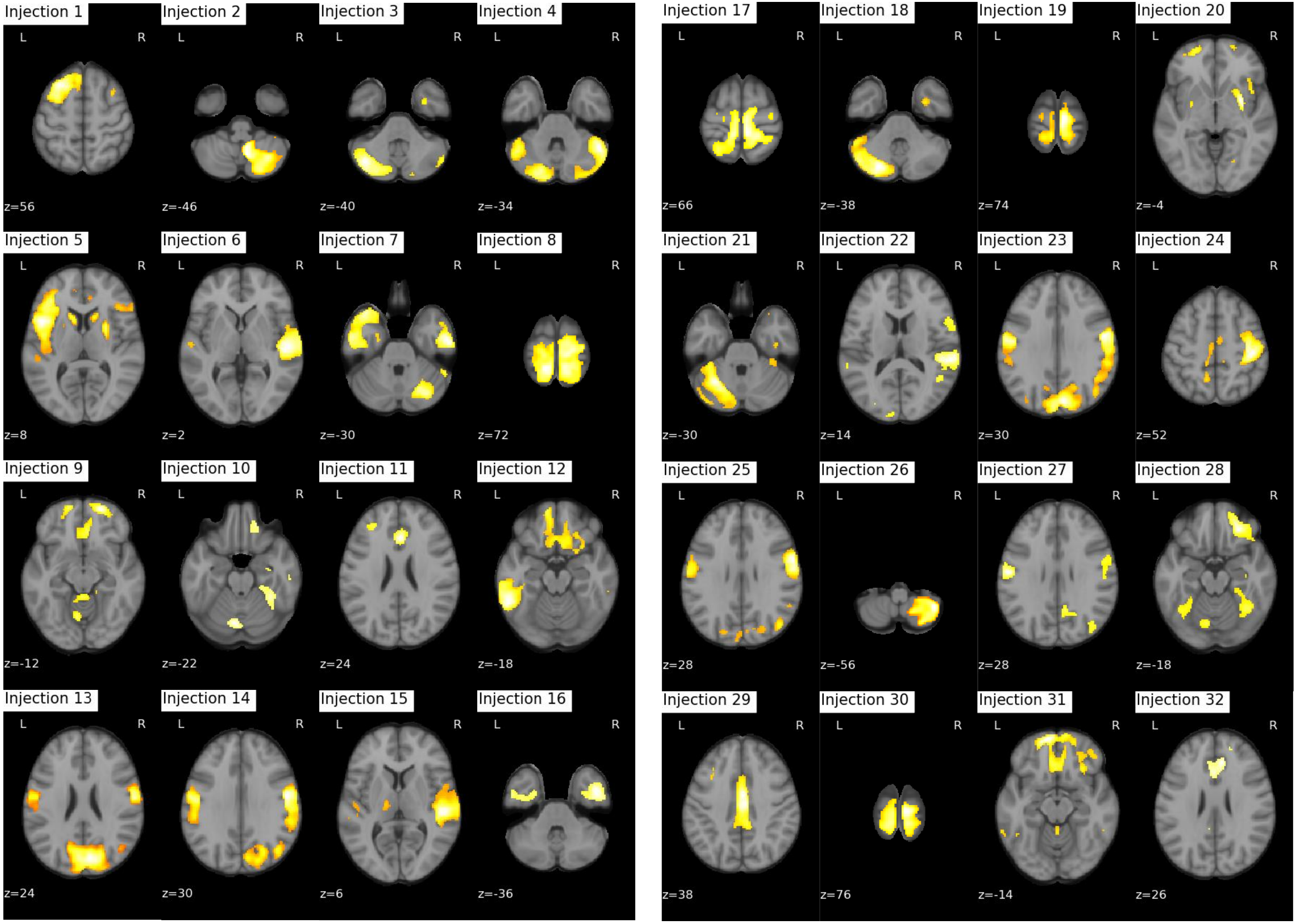}
\caption{Visualization of the 32 injection positions identified in our study, which are reprojected onto brain space. Each subfigure represents an injection map, with values ranging from 0 to 1 that indicate the probability of signal injection at various brain locations. A threshold value of 0.6 has been applied to enhance clarity.}
\end{figure}

To further delineate the distribution of these injection maps, Figure 9 overlays the contours of all injection maps, with each map differentiated by a unique color. Additionally, to investigate the neurological significance of these injection signals, we conducted a meta-analysis using the Automated Anatomical Labeling (AAL) atlas (version 3, comprising 170 predefined brain regions)\cite{rolls2020automated}. In this analysis, we calculated the overlap rate between the AAL regions and each of the injection maps. The resulting overlap heatmap is displayed in Figure 10 (top), accompanied by a summary of the top five overlapping AAL regions for each map in Supplementary Table S1. From our results, we observed that all of these injection maps exhibit significant neurological relevance. For example, injection map \#1 reveals concurrent activation in the Middle Frontal Gyrus (MFG, area 5), Inferior Temporal Gyrus (ITG, areas 93 and 94), Superior Frontal Gyrus (SFG, area 3), and Medial Superior Frontal Gyrus (SFG, area 19). These activations are indicative of various cognitive functions: the MFG is associated with executive functions, the ITG with visual processing and object recognition, the SFG with higher cognitive functions and motor control, and the SFG with executive control and self-referential thoughts. These results emphasize the intricate interplay among visual processing, cognitive control, and motor planning, highlighting the capacity of injection map \#1 to integrate diverse cognitive functions in response to specific stimuli or tasks.

\begin{figure}[htb]
\centering
\includegraphics[width=\linewidth]{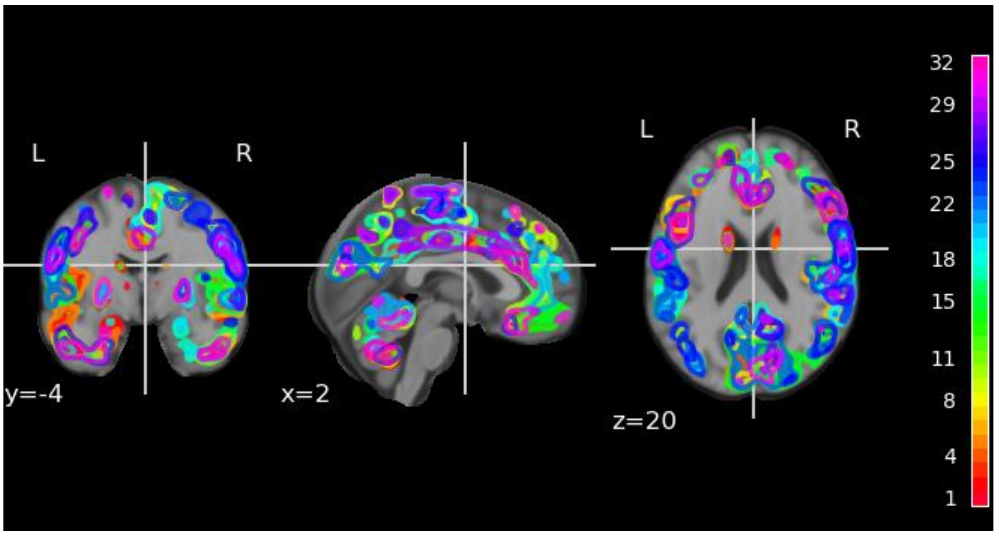}
\caption{An overlay of injection maps depicting their spatial distribution throughout the brain, with each map uniquely color-coded for clear differentiation.}
\end{figure}

\begin{figure}[htb]
\centering
\includegraphics[width=\linewidth]{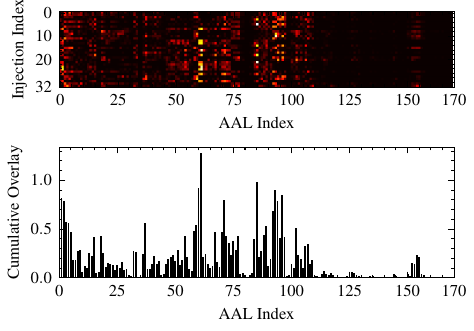}
\caption{Top: the overlap heatmap of injection maps with AAL atlas regions. Bottom: cumulative overlap rate for each AAL region, highlighting their importance in the injection signal.}
\end{figure}

In addition to analyzing individual injection maps, we also determined which brain areas within the AAL atlas are most frequently targeted in our study. To do this, we calculated the cumulative overlap rate for each AAL region, where a higher total value indicates a greater likelihood of receiving injection signals. The results are displayed in Figure 10 (bottom), with details on the top ten regions provided in Supplementary Table S1. Our analysis reveals that the injection signals provide profound insights into the complex interactions among sensory processing, motor functions, and cognitive activities. Notably, the bilateral activation observed in the Postcentral Gyrus (PoCG) highlights its crucial role in integrating somatosensory information. The activity in the right Superior Temporal Gyrus (STG) suggests its involvement in auditory processing and potentially language comprehension. The engagement of the Inferior Temporal Gyrus (ITG) across both hemispheres points to a significant involvement in complex visual processing, particularly in object recognition. Furthermore, the activation of the Precentral Gyrus (PreCG) is consistent with its recognized function in motor execution and planning. The Precuneus (PCUN) activation indicates its integration of visuospatial information and episodic memory, shedding light on its role in self-related cognitive processes. The activation in cerebellar regions (Cerebellar Crus 1 and 2) extends beyond traditional motor functions, hinting at their involvement in cognitive activities. The activation of the Superior Frontal Gyrus (SFG) is aligned with its contribution to higher cognitive functions, suggesting a network of regions collaboratively supporting a wide array of cognitive and motor tasks. Overall, these injection maps underscore the intricate interplay between various brain regions in supporting a spectrum of neural functions, from sensory integration to cognitive processing.

Following our examination of the role of injection positions, we proceeded to investigate the fluctuations of injection signals in a subsequent experiment. Unlike static positions, these fluctuations are the sole factors that dynamically change in response to transitions in brain states. Figure 11 provides a visualization of these fluctuations for two randomly selected subjects (IDs 100206 and 100307) across four distinct tasks. Our results show that all injection signals tend to oscillate around a zero value, which demonstrates the model's stability at the system level. Furthermore, we observed that these fluctuations vary across subjects, even when they are engaged in the same task. For instance, in Figure 11, subject 100206 (represented by the red curve) displays significant variation compared to subject 100307 (blue curve). Significantly, our analysis revealed that the resting state consistently exhibits the most notable variability among subjects. To illustrate, we calculated the L2 distance between the signal fluctuations of subjects 100206 and 100307. The results presented in Figure 12 show that the L2 distance in the resting state is nearly twice as large as that observed during other task conditions, underscoring the unique nature of resting-state brain activity. We believe that these variations reflect individual differences in brain functional activity. Despite the variability in injection signal fluctuations among subjects, in the following section, we will demonstrate how these signals can be effectively used to detect consistent task-related brain activities across different individuals.

\begin{figure}[htb]
    \centering
  \subfloat[EMOTION]{%
       \includegraphics[width=0.5\linewidth]{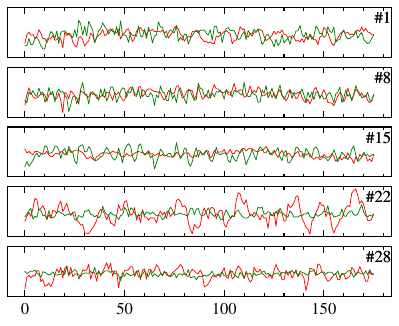}}
    \hfill
  \subfloat[LANGUAGE]{%
        \includegraphics[width=0.5\linewidth]{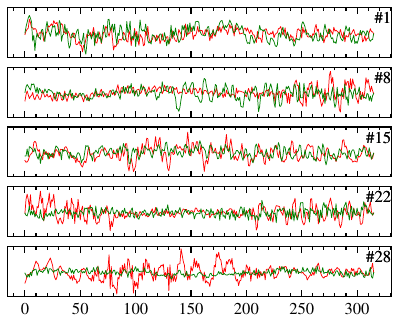}}
    \\
  \subfloat[RELATIONAL]{%
        \includegraphics[width=0.5\linewidth]{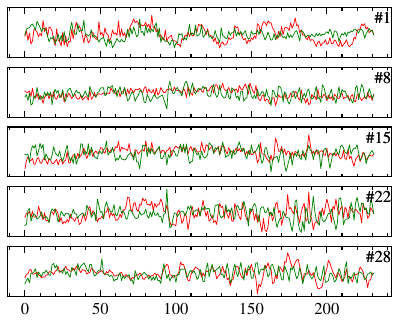}}
    \hfill
  \subfloat[REST]{%
        \includegraphics[width=0.5\linewidth]{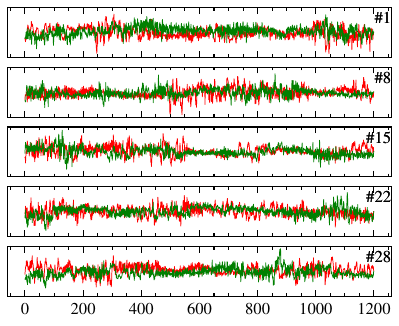}}
  \caption{Visualization of signal fluctuation patterns across four different tasks for two randomly selected subjects.}
  \label{fig1} 
\end{figure}

\begin{figure}[htb]
\centering
\includegraphics[width=\linewidth]{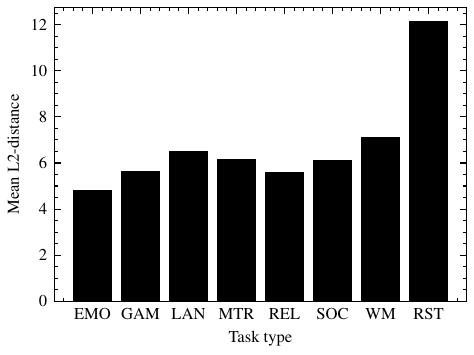}
\caption{Average L2 distance of signal fluctuations between randomly selected subjects.}
\end{figure}

\subsection{Exploring Brain Activities through Temporal Queries}

In the first scenario, we only have precise knowledge of when the event occurs. To understand the content of these events, constructing temporal queries is necessary, as detailed in the methods section. This scenario is common in many studies. For instance, we might know exactly when a subject watches a movie, but determining the specific content of the movie or the subject's thoughts about it remains a challenge.

An example of constructed temporal queries is shown in Figure 13, involving motor tasks with six events. In this figure, the magenta curve illustrates the absence of events over time, while the purple curve represents the constructed query. Unlike GLM, this study does not directly apply regression analysis to voxels to identify brain areas. Instead, after initializing with injection signals, we treat the BDI model as the subject's brain. Our aim is to find spatial maps that allow the model's output responses to align with the temporal queries. These maps are considered representations of brain areas activated by corresponding events. The search process is thoroughly detailed in the methods section. Briefly, during this process, we freeze the model weights and injection signals and apply backpropagation to tune the position vectors until the model's response matches the temporal query. The resulting position vector then can be remapped to brain space, and serves as the spatial maps that we use for further analysis.

\begin{figure}[htb]
\centering
\includegraphics[width=\linewidth]{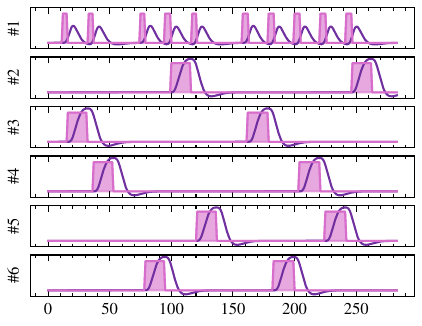}
\caption{An example of a constructed temporal query for a Motor task, which includes six events.}
\end{figure}

\begin{figure}[htb]
\centering
\includegraphics[width=\linewidth]{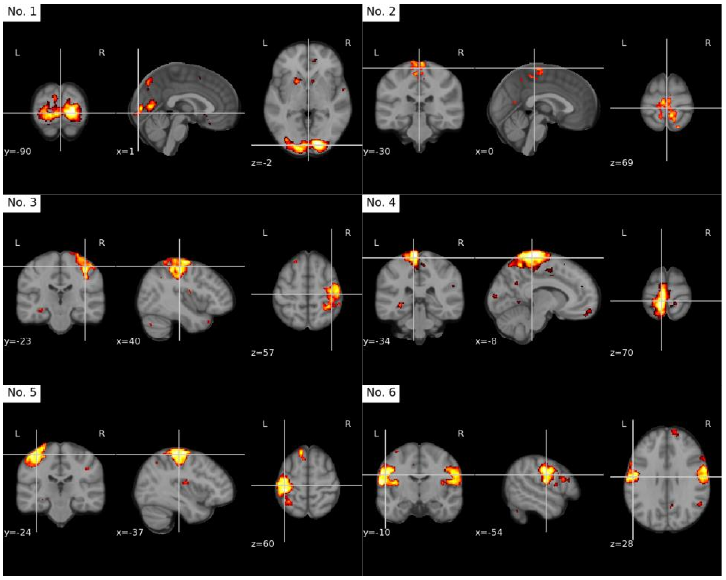}
\caption{Position vectors that identified by the motor temporal query, each mapped into brain space for visualization. A threshold of 0.6 is applied.}
\end{figure}

The performance of temporal query experiments is evaluated in two aspects. Firstly, we verified the accuracy of identifying event-related brain areas. To measure this accuracy, we used the results from General Linear Models (GLM) as a benchmark and calculated the overlap rate between our identified spatial maps and those from GLM. A higher overlap rate indicates better accuracy. Figure 14 displays spatial maps from six motor task events, with the corresponding overlap rates detailed in Table I (first row). For comparison, Table I also features results from previous studies utilizing Independent Component Analysis (ICA)\cite{svensen2002ica}, Temporal Restricted Boltzmann Machine (tRBM)\cite{hjelm2014restricted}, and Spatial Restricted Boltzmann Machine (sRBM)\cite{hu2018latent}. Our model outperformed these methods, achieving the highest accuracy in detecting brain activities for all six events. For the other tasks, the overlap rates are summarized in Table II, which indicates that the average overlap rates of our model are twice as high as those achieved with the sRBM.

\begin{table}[H]
\begin{center}
\caption{Overlap of spatial maps with GLM in motor task. ICA, tRBM, and sRBM are listed for comparison.}
\label{tab1}
\begin{tabular}{lcccccc}
\toprule  
  & E1 &E2 &E3& E4 & E5 & E6\\
\midrule  
BDI-GLM&\textbf{0.55} &\textbf{0.61} &\textbf{0.58} & \textbf{0.50}  &\textbf{0.61} & \textbf{0.62}\\ 
sRMB-GLM&0.27 &0.49 &0.53 & 0.37  &0.41 & 0.52\\ 
tRMB-GLM&0.13 &0.22 &0.06 & 0.14  &0.48 & 0.31\\
ICA-GLM&0.19 &0.33 &0.35 & 0.27  &0.31 & 0.51\\
\bottomrule 
\end{tabular}
\end{center}
\end{table}

\begin{table}[H]
\begin{center}
\caption{Average overlap of spatal maps with GLM in all tasks. ICA, tRBM, and sRBM are listed for comparison.}
\label{tab1}
\begin{tabular}{lcccc}
\toprule  
 Tasks & BDI-GLM &sRMB-GLM &tRMB-GLM&ICA-GLM\\
\midrule  
GAMBLING & \textbf{0.733} &0.385 &0.16  &  0.335\\
EMOTION & \textbf{0.839} &0.35 & 0.11 & 0.475\\
LANGUAGE & \textbf{0.62} &0.32 & 0.02 & 0.14\\
MOTOR &  \textbf{0.58} &0.432 & 0.22 & 0.33\\
RELATIONAL & \textbf{0.799} &0.325 & 0.25 & 0.315\\
SOCIAL & \textbf{0.635} &0.3 & 0.11 & 0.095\\
WM & \textbf{0.61} &0.36 & 0.062 & 0.0125\\
\bottomrule 
\end{tabular}
\end{center}
\end{table}

In the second validation, after obtaining the spatial maps, we re-input them into the model and calculated the Pearson correlation between the temporal responses and the temporal queries across all testing subjects. Figure 15 displays the average temporal response and the corresponding temporal query for motor tasks, with their correlation values summarized in Table III. The correlation values for other tasks are summarized in Table IV. Similar to the spatial maps, we achieved high correlation values, which further verified the accuracy of the spatial maps identified through temporal queries.

\begin{figure}[htb]
\centering
\includegraphics[width=\linewidth]{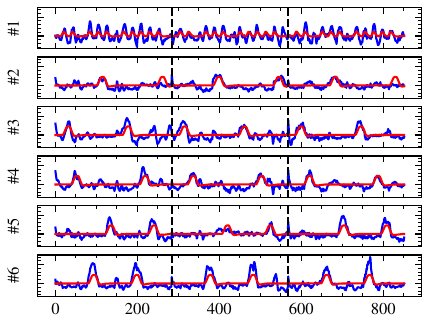}
\caption{The temporal response of the model for each of the six events, shown here for three subjects. The blue line represents the temporal query, and the red line illustrates the model's response}
\end{figure}

Once the spatial maps have been identified, we can perform a meta-analysis on these maps to indirectly determine the content of the events. Overall, by utilizing temporal queries with the proposed model in this study, we can efficiently identify brain areas highly related to specific events, based solely on their timing. In subsequent experimental results, we will continue to focus on motor tasks, as these have been extensively studied in previous research, allowing for a robust comparison. For additional results on other tasks, please refer to the supplementary material.

\begin{table}[H]
\begin{center}
\caption{Average Pearson correlation values across all subjects between the temporal query and model response in each events of motor task.}
\label{tab1}
\begin{tabular}{lc}
\toprule  
Events& Average Pearson Correlation\\
\midrule  
\#1& 0.786\\ 
\#2& 0.616\\
\#3& 0.761\\
\#4& 0.742\\
\#5& 0.783\\
\#6& 0.841\\
\bottomrule 
\end{tabular}
\end{center}
\end{table}

\begin{table}[H]
\begin{center}
\caption{Average Pearson correlation values across all subjects between the temporal query and model response in each task.}
\label{tab1}
\begin{tabular}{lcc}
\toprule  
Tasks & Average Pearson Correlation \\
\midrule  
GAMBLING & 0.682\\
EMOTION & 0.775\\
LANGUAGE & 0.651\\
MOTOR & 0.755\\
RELATIONAL & 0.665\\
SOCIAL & 0.721\\
WM & 0.62\\
\bottomrule 
\end{tabular}
\end{center}
\end{table}

\subsection{Determining Event Timing with Brain Spatial Queries}

In another scenario where the contents of the events are known, specifically, brain activation related to a task has been previously identified, but the precise timing of the tasks remains unclear. We can utilize the known brain maps as a spatial query to determine the exact timing of the events. This query process is quite straightforward and requires no additional tuning. As previously discussed, we simply input the spatial query into the model, which then generates a temporal response. By examining and comparing the relative strength of responses at each time point, we can determine the timing of the events.

From our experiments, the high correlation value between the temporal response and the task design presented in Table IV validates the performance of our method for determining event timing. However, it is important to note that obtaining accurate spatial maps of an event in advance is often challenging. The brain is a dynamic and highly energy-efficient system where a single brain map may be involved in multiple tasks. If we liken a brain map to a ``word" in natural language, then a task can be considered a ``sentence" composed of many words. Consequently, identifying all the brain maps associated with a task is quite challenging. Missing any of the ``words" can potentially lead to unpredictable results.

To identify all brain maps associated with a task, we proposed three potential methods in this study. The first involves using a temporal query, as previously discussed. Once the timing of an event is known, we can detect all related brain areas without omission. The second method employs a meta-analysis approach to manually construct a spatial query. For example, we can use the Automated Anatomical Labeling (AAL) atlas for this purpose. The AAL atlas divides the brain into predefined regions based on anatomical landmarks, with each area being assigned a unique functional description. Figure S4 (see Supplementary Materials) illustrates the use of the AAL atlas to construct three spatial queries for motor events involving visual cues and finger tapping (left or right). We selected three AAL brain areas, identified as \#57, \#73, and \#74, based on our prior knowledge of their functions to serve as spatial queries. Their corresponding temporal responses are depicted in supplementary materials (Figure S5). From our experiment, we achieved reasonably good correlation values with the task design, with correlations of 0.527, 0.47, and 0.556, respectively.

Additionally, the predefined AAL atlas can also be used for volume-wise classification of brain states. For instance, in supplementary materials (Figure S6), we utilized all 170 AAL brain areas as a sequence of spatial queries. After obtaining their corresponding temporal responses, we applied a SoftMax activation to normalize these responses between 0 and 1. This allows us to infer the subject's brain state at each time point by identifying the area with the strongest response. Although manually designing spatial queries is technically feasible, it requires extensive work in labeling the functions of each area and can easily lead to ambiguities. In the next section, we will introduce a third method for constructing spatial queries, which is better suited to modern research requirements.

\subsection{Dialoguing with the Brain Through Text and Image Inputs}

In the previous section, we validated the performance of our model in enabling dialogue with the brain through temporal or spatial queries to ascertain the experiences of a subject. However, constructing these queries requires some neurological expertise, which may pose difficulties for researchers from other fields. Although we previously introduced two methods for creating spatial queries, there remains a need for a more user-friendly and widely accessible approach. To meet this requirement, we propose a third method here that could translates text (English) or visual images into spatial queries. For more details on this method, please refer to the Methods section. Briefly, our study utilizes a renowned multimodal model known as Contrastive Language–Image Pre-training (CLIP)\cite{radford2021learning}. By integrating it with our Brain-Dialogue Interface (BDI) model, we can convert text or image inputs into spatial queries, thereby simplifying the process of communicating with the brain.

In the experiment, we evaluated the accuracy of translating text and images into spatial queries by measuring their overlap rate with ground truth and the correlation of temporal responses with the underlying task design. We will first discuss the experimental results of the text-based spatial queries and then explore the outcomes for image-based queries.

\begin{figure}[htb]
\centering
\includegraphics[width=\linewidth]{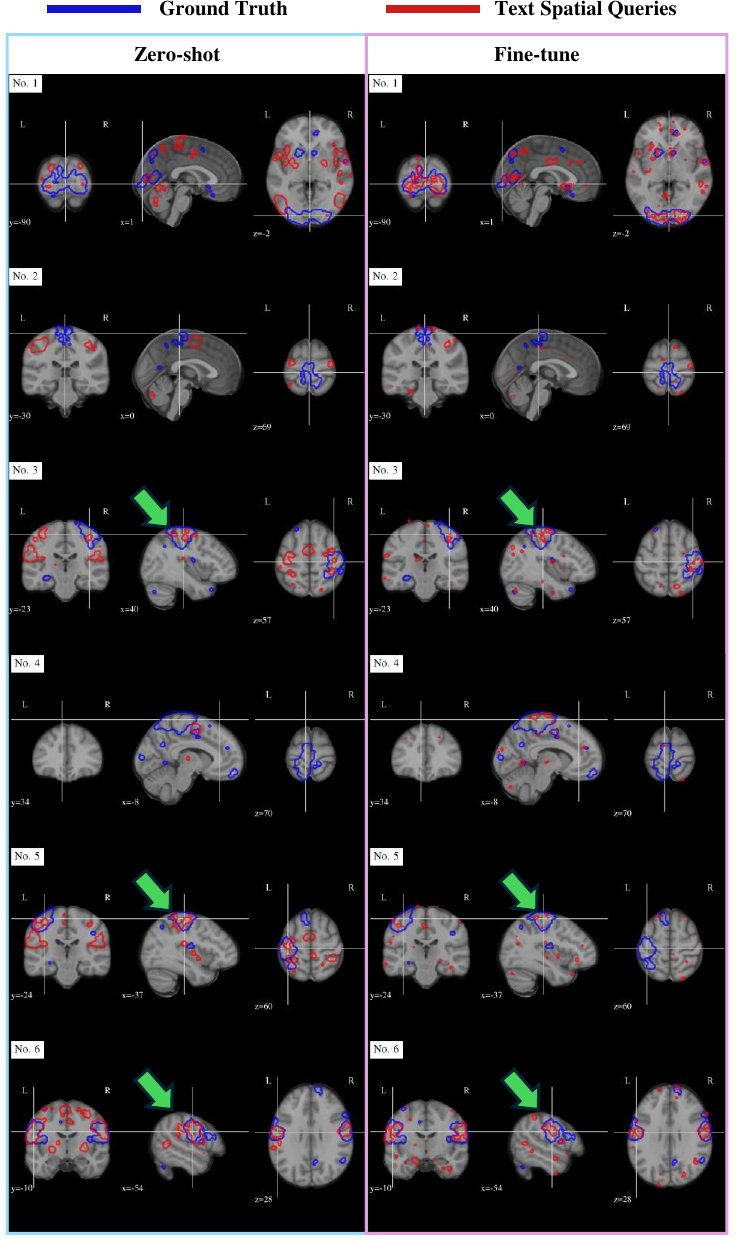}
\caption{Visualization of text spatial queries (highlighted with red contours) in six events of motor task, with ground-truth  (blue contours) overlapped for comparison. The left panel presents results from zero-shot evaluation, while the right panel illustrates results from fine-tune experiments.}
\end{figure}

Figure 16 presents examples of spatial queries derived from text inputs related to six motor task events, which we refer to simply as `text spatial queries' for clarity. In this figure, red contours highlight the brain regions targeted by text queries, while blue contours represent the ground truth of the corresponding task events. We used spatial maps from temporal queries described in the previous section as the ground truth for evaluation. Furthermore, Figure 16 includes a left panel displaying results from `zero-shot' experiments, in which the text spatial queries are generated without any prior information about the tasks. The right panel depicts `fine-tuned' results, where information from the task design is used to fine-tune the parameters of the spatial query layer.

\begin{figure}[htb]
\centering
\includegraphics[width=\linewidth]{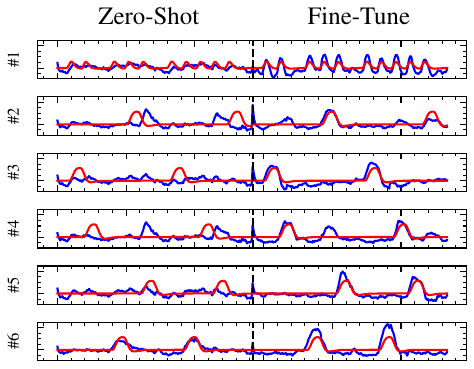}
\caption{Average temporal response of the BDI model to text queries. The left panel shows results from zero-shot evaluations, while the right panel presents outcomes from fine-tuning experiments.}
\end{figure}

From the results, we can see that even though the model was not provided with task information in the 'zero-shot' experiments, it was still capable of translating text descriptions of tasks into relevant spatial queries. For instance, events 3 (squeeze left toes), 5 (squeeze right toes), and 6 (move tongue), which are marked by green arrows in the figure, accurately identify brain regions within the corresponding motor areas, consistent with neuroscience findings. Similar results can be drawn from other tasks. However, we also observe that these spatial queries encompass some event-unrelated areas, which affect the temporal response of the model. This phenomenon is evident in Figure 17, which displays the temporal responses corresponding to each event. Notably, apart from event 6 (move tongue), the temporal responses from other text spatial queries show a relatively low correlation with the task design. Compared to the `zero-shot' approach, the performance of text translation specifically in terms of spatial query accuracy and temporal response correlation can be significantly improved through `fine-tuning'. The visualized results for the six fine-tuned motor task events are illustrated in Figures 16 and 17. The overlap rates and temporal response correlations for all text-based task queries are comprehensively summarized in Tables VI, VII, VIII, and IX.

In addition to evaluating the overlap rate and Pearson correlation of text queries, we also conducted a brain state classification experiment. In this experiment, our aim was to classify brain states at each time point by assigning a state label according to the model’s temporal response to each text query. The results are displayed in Table V. From these results, we observed that our method could perform zero-shot classification for some tasks, for instance, we achieved accuracies of 68.27\%, 15.46\%, and 21.91\% in Emotion, Language, and Motor tasks respectively. Given the complexity of human brain states, it is quite interesting that we achieved these results without any pretraining. Moreover, the classification accuracy was significantly enhanced when task design information was incorporated, yielding good accuracy across all tasks (Fine-tune in Table VII). This experiment further substantiates the capability of using text inputs to retrieve information from the brain.

\begin{table}[H]
\begin{center}
\caption{Decoding accuracy (\%) evaluated at each time point for various tasks using text queries as input}
\begin{tabular}{lcccc}
\toprule  
Tasks & ZeroShot & Finetune  & Chance\\
\midrule  
EMOTION & 68.27& 81.01  & 50\\ 
GAMBLING & 21.03 & 27.89  & 20\\
LANGUAGE & 15.46 & 24.10 & 11.11\\
MOTOR & 21.91 & 57.47& 16.7\\
RELATIONAL & 34.22 & 45.11  & 33.3\\
SOCIAL & 25.72 & 40.36  & 25\\
WM &6.14 & 22.65  & 6.25\\
\bottomrule 
\end{tabular}
\end{center}
\end{table}

\begin{figure}[htb]
\centering
\includegraphics[width=\linewidth]{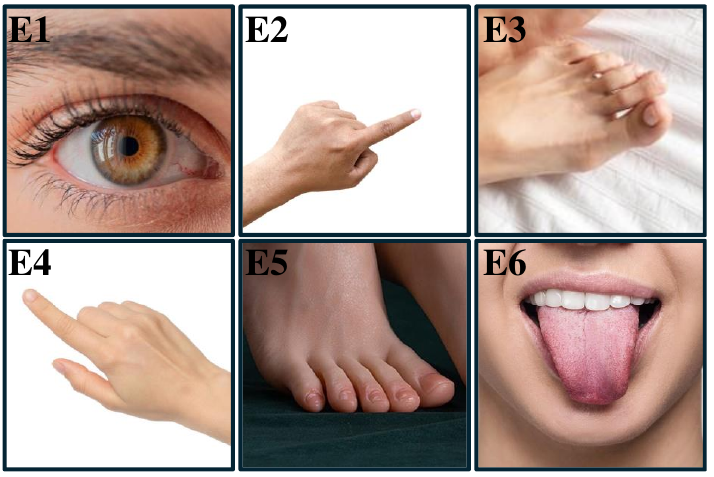}
\caption{Images randomly selected from the Internet, used as image queries for six events in motor task.}
\end{figure}

Given that CLIP is a multimodal model that has achieved good alignment between text and images, we should technically be able to translate images into spatial queries as easily as we do with text. However, our experiments revealed significant challenges in finding appropriate images that correspond to specific brain states. For instance, it is difficult to use an image to represent the brain state associated with `working memory' or other abstract concepts like `interaction' or `movement.' Despite these challenges, we conducted a preliminary experiment to assess the feasibility of translating images into spatial queries. As shown in Figure 18, we selected six images from the internet that potentially reflect the content of six motor task events. Since images and text can be projected into the same feature space, using a process similar to text queries, these images could also be translated into spatial queries. The resulting spatial maps are illustrated in Figure 19. Similar to our findings with text queries, we can also obtain event-related spatial maps using image queries. However, we observed that, on average, image queries tend to engage more unrelated active brain areas compared to text queries, which is likely due to the ambiguous nature of images. As a result, except for event 1 (visual cue), which achieved a 0.424 correlation with the task design, the correlations for other events in image queries are relatively lower at the group level (Table VII). In future work, we plan to experiment with video as input, which may more precisely capture the intentions behind the queries.

\begin{figure}[htb]
\centering
\includegraphics[width=\linewidth]{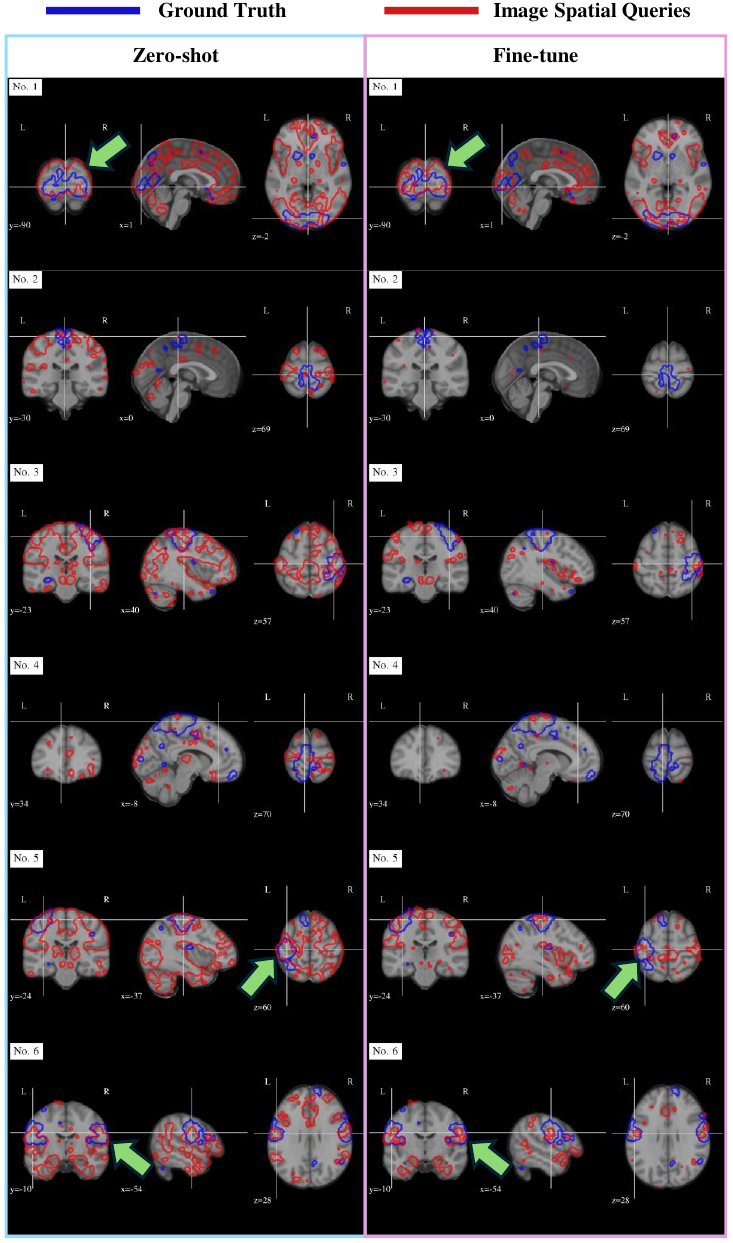}
\caption{Visualization of image spatial queries (highlighted with red contours) in six events of motor task, with ground-truth  (blue contours) overlapped for comparison. The left panel presents results from zero-shot evaluation, while the right panel illustrates results from fine-tune experiments.}
\end{figure}

\begin{table}[H]
\begin{center}
\caption{Overlap of text or image queries with ground truth in motor task. `T-ZeroShot' and `T-Finetune' denote text zero shot and finetune experiments, respectively, while `I-ZeroShot' and `I-Finetune' represent image zero shot and finetune experiments.}
\label{tab1}
\begin{tabular}{lcccc}
\toprule  
Events & T-ZeroShot & T-Finetune  &  I-ZeroShot & I-Finetune\\
\midrule  
E1& 0.096 & 0.715 & 0.375 & 0.311\\
E2& 0.038 & 0.623 & 0.108 & 0.457\\
E3& 0.118 & 0.674 & 0.298 & 0.304\\
E4& 0.069 & 0.709 & 0.119 & 0.456\\
E5& 0.232 & 0.666 & 0.379 & 0.304\\
E6& 0.243 & 0.691 & 0.257 & 0.34\\
\bottomrule 
\end{tabular}
\end{center}
\end{table}

\begin{table}[H]
\begin{center}
\caption{Average correlations between model temporal responses and task designs in the motor task using text or image queries. `T-ZeroShot' and `T-Finetune' refer to text zero shot and finetune experiments respectively. `I-ZeroShot' and `I-Finetune' represent image zero shot and finetune experiments.}
\label{tab1}
\begin{tabular}{lcccc}
\toprule  
Events & T-ZeroShot & T-Finetune  &  I-ZeroShot & I-Finetune\\
\midrule  
E1& 0.245 & 0.717 & 0.359 & 0.424\\
E2& -0.167 & 0.514 & 0.072 & 0.13\\
E3& -0.055 & 0.708 & -0.01 & 0.007\\
E4& -0.062 & 0.65 & -0.01 & -0.011\\
E5& 0.121 & 0.757 & 0.02 & 0.042\\
E6& 0.572 & 0.801 & 0.053 & 0.167\\
\bottomrule 
\end{tabular}
\end{center}
\end{table}

\begin{table}[H]
\begin{center}
\caption{Average overlap of text and image queries with ground truth in all task. `T-ZeroShot' and `T-Finetune' denote text zero shot and finetune experiments, respectively, while `I-ZeroShot' and `I-Finetune' represent image zero shot and finetune experiments.}
\begin{tabular}{lcccc}
\toprule  
Tasks & T-ZeroShot & T-Finetune  &  I-ZeroShot & I-Finetune \\
\midrule  
EMOTION & 0.289 & 0.632 & - & - \\ 
GAMBLING & 0.069 & 0.599 & - & -\\
LANGUAGE & 0.164 & 0.658 & - & -\\
MOTOR & 0.132 & 0.68 & 0.256 & 0.362\\
RELATIONAL & 0.113 & 0.58 & - & -\\
SOCIAL & 0.159 & 0.647 & - & -\\
WM &0.08 & 0.64 & - & -\\
\bottomrule 
\end{tabular}
\end{center}
\end{table}

\begin{table}[H]
\begin{center}
\caption{Average correlations between model temporal responses and task designs in all task using text or image queries. `T-ZeroShot' and `T-Finetune' refer to text zero shot and finetune experiments respectively. `I-ZeroShot' and `I-Finetune' represent image zero shot and finetune experiments.}
\begin{tabular}{lcccc}
\toprule  
Tasks & T-ZeroShot & T-Finetune  &  I-ZeroShot & I-Finetune \\
\midrule  
EMOTION & 0.279 & 0.727 & - & - \\ 
GAMBLING & -0.014 & 0.488 & - & -\\
LANGUAGE & 0.245 & 0.515 & - & -\\
MOTOR & 0.109 & 0.691 & 0.079 & 0.126\\
RELATIONAL & 0.331 & 0.499 & - & -\\
SOCIAL & 0.138 & 0.609 & - & -\\
WM &0.039 & 0.445 & - & -\\
\bottomrule 
\end{tabular}
\end{center}
\end{table}

\section{Conclusion}
In conclusion, our study introduces an innovative brain decoding model that surpasses the constraints of traditional static methods by utilizing a dynamic, unsupervised learning approach. This model, which strives to accurately reflect brain activity and facilitate interactive dialogues, offers a promising addition to the field of brain decoding technology. The combination of our decoding model with sophisticated textual input systems introduces exciting new prospects for user interaction, enabling more natural and adaptable control over brain decoding applications. Our experiments suggest the potential effectiveness of dynamic, unsupervised decoding in unraveling complex neural activity. Looking ahead, we expect that ongoing enhancements and applications of such models will profoundly deepen our understanding of brain functions.

\bibliographystyle{IEEEtran}
\bibliography{ref}

\newpage
\vfill

\end{document}